\documentclass[aps,prb,twocolumn,superscriptaddress,showpacs]{revtex4}
\usepackage{feyn}
\usepackage{graphicx}
\usepackage{latexsym}
\usepackage{amssymb}
\usepackage{amsmath}
\usepackage{amsfonts}
\usepackage{bm}
\usepackage{multirow}
\usepackage{color}
\usepackage{comment}
\newcommand{\ii}{\mathrm{i}}

\newcommand{\SO}{\mathrm{SO}}

\newcommand{\U}{\mathrm{U}}

\newcommand{\figref}[1]{Fig.\,\ref{#1}}

\newcommand{\beq}{\begin{equation}}
\newcommand{\eeq}{\end{equation}}
\newcommand{\beqn}{\begin{eqnarray}}
\newcommand{\eeqn}{\end{eqnarray}}

\DeclareMathAlphabet{\mathbbold}{U}{bbold}{m}{n}

\def\sgn{{\rm sgn}}

\def\U{{\rm U}}

\begin{document}

\title{Green's function Zero and Symmetric Mass Generation}

\author{Yichen Xu}
\affiliation{Department of Physics, University of California,
Santa Barbara, CA 93106, USA}

\author{Cenke Xu}
\affiliation{Department of Physics, University of California,
Santa Barbara, CA 93106, USA}

\begin{abstract}

It is known that, under short-range interactions many topological
superconductors (TSC) and topological insulators (TI) are
trivialized, which means the boundary state of the system can be
trivially gapped out by interaction without leading to symmetry
breaking or topological ground state degeneracy. This phenomenon
is also referred to as ``symmetric mass generation" (SMG), and has
attracted broad attentions from both the condensed matter and high
energy physics communities. However, after the trivialization
caused by interaction, some trace of the nontrivial topology of
the system still persists. Previous studies have indicated that
interacting topological TSC and TI could be related to the ``zero"
of Green's function, namely the fermion Green's function $G(\ii
\omega \rightarrow 0) = 0$. In this work, through the general
``decorated defect" construction of symmetry protected topological
(SPT) states, we demonstrate the existence of Green's function
zero after SMG, by mapping the evaluation of the Green's function
to the problem of a single particle path integral. This method can
be extended to the cases without spatial translation symmetry,
where the momentum space which hosts many quantized topological
numbers is no longer meaningful. Using the same method one can
demonstrate the existence of the Green's function zero at the
``avoided topological transition" in the bulk of the system.

\end{abstract}

\maketitle

\section{Introduction}


Short range interactions can modify the classification of
topological superconductors (TSC) and topological insulators (TI)
in the classic ``ten-fold way" table for free
electrons~\cite{ludwigclass1,ludwigclass2,kitaevclass}. The most
prominent feature of a TSC or a TI is at its boundary, i.e in the
noninteracting limit, the boundary of a TSC or TI should be
gapless unless the boundary breaks the defining symmetry of the
system. A short range interaction can enrich the phenomena at the
boundary of a TSC and TI: it can drive the boundary into a
spontaneous symmetry breaking phase, or a gapped topological phase
which preserves all the
symmetries~\cite{TOQi,TOSenthil,TOAshvin,TOMax}. But it has also
been realized that, a short range interaction may trivialize some
of the TSCs and TIs, in the sense that short range interaction can
``trivially" gap out the boundary of some TSCs and TIs without
breaking any symmetry or leading to any ground state degeneracy.
The first $1d$ example of this interaction-trivialized TSC was
found in Ref.~\onlinecite{fidkowski1,fidkowski2}, and soon other
examples were found in all
dimensions~\cite{qiz8,yaoz8,zhangz8,TOAshvin,senthilhe3,youxu2,fuz4,hermelez4,z42,wenz4},
For example, now it is known that 16 copies of the TSC $^3$He-B
phase is trivialized by interaction, hence although this TSC in
the noninteracting limit has a $\mathbb{Z}$ classification, under
interaction $^3$He-B has a $\mathbb{Z}_{16}$
classification~\cite{TOAshvin,senthilhe3}.

Interaction trivialized TSC and TI has a deep relation with
another phenomenon called ``symmetric mass generation" (SMG). In
the noninteracting limit the boundary of a $(d+1)$-dimensional TSC
is described by a $d$-dimensional gapless Majorana fermion (or
chiral Majorana fermion depending on the dimensionality), which
carries with it certain 't Hooft anomaly of the defining
symmetries of the TSC. A mass term of the boundary fermion will
explicitly break the symmetry, and hence is prohibited to exist.
When interaction reduces the classification of a TSC from
$\mathbb{Z}$ to $\mathbb{Z}_N$, it means that for $N$ copies of
the $d$-dimensional Majorana fermions, it is possible to generate
a gap through interaction without any degeneracy at the
$d$-dimensional boundary, and the expectation value of any fermion
bilinear mass operator is zero. The mechanism of SMG is in stark
contrast with the ordinary mass generation of a Dirac or Majorana
fermion (the well-known Gross-Neveu-Yukawa-Higgs
mechanism~\cite{gross}), which is caused by the condensation of a
boson that couples to the fermion bilinear mass term. The
condensation of the boson will break the symmetry of the system,
and lead to a nonzero expectation value of a fermion mass term.
The SMG has attracted broad attentions from both the condensed
matter~\cite{kevinQSH,he2016,smg1,smg2} and high energy
communities~\cite{hepsmg1,hepsmg2,hepsmg3,hepsmg4,hepsmg5,hepsmg6,hepsmg7,hepsmg8}
in the last few years, partly motivated by the observation that
the SMG mechanism may be related to the lattice regularization of
chiral gauge theories such as the Grand Unified
Theories~\cite{wengut1,wengut2,wengut3,xugut1,xugut2,wengut5,wengut4,tonggut}.

A natural question one may ask is that, after the interaction
``trivializes" the system, or after the SMG, is there still any
remaining trace of the nontrivial topology of the original
noninteracting system? Or for a system with a fully gapped
spectrum, how do we know the gap originates from the mechanism of
SMG? The two most important features of TSCs or TIs are their
stable boundary states, and the unavoidable bulk topological phase
transition from the trivial insulator. When a TSC or TI is
trivialized by interaction, both features mentioned above no
longer robustly hold. It has been proposed before that strongly
interacting TIs and TSCs may have a close relation with the zero
of fermion Green's
functions~\cite{gurarie1,gurarie2,kevinQSH,balentszero,youzero,hepsmg5,smg1,smg2}.
In this work we use the general ``decorated defect" construction
of symmetry protected topological (SPT)
states~\cite{chenluashvin}, and map the computation of the fermion
Green's function to a problem of single particle path integral.
Our method demonstrates in arbitrary dimensions the existence of
fermion Green's function zero, after the interaction trivializes
the TSC and TI, i.e. after the symmetric mass generation. One of
the previous arguments (which will be reviewed later) for the
existence of the Green's function zero relies on the quantized
topological number defined with fermion Green's function in the
momentum space. Our method can be generalized to the cases without
translation symmetry, where a momentum space is no longer
meaningful.

\section{Green's function ``zero" from decorated defects}


Intuitively the decorated defect construction of a SPT state
follows three steps: (1) one starts with a bosonic system with
certain symmetry $G$, and drive the bosonic system into an ordered
state with spontaneous symmetry breaking of the symmetry $G$,
which allows topological defects; (2) decorate the topological
defects with a lower dimensional SPT state, and (3) eventually
proliferate the defects to restore the symmetry $G$. This
decorated defect construction was originally designed for bosonic
SPT states~\cite{chenluashvin}, but it also applies to fermionic
TSCs and TIs. For example, the $2d$ TSC with the $Z_2 \times
Z_2^T$ symmetry ($p \pm \ii p$ TSC) can be constructed by
decorating the $Z_2$ domain wall with a $1d$ TSC with the
time-reversal ($Z_2^T$) symmetry (the $1d$ BDI class TSC or the so
called Kitaev's chain); The $3d$ TSC (or TI) of the AIII class can
be constructed by decorating a $\U(1)$ vortex line with the
Kitaev's chain.

We will start with the example of $2d$ TSC with $Z_2 \times Z_2^T$
symmetry, and first compute the fermion Green's function at the
$1d$ boundary of the system. Consider a $2d$ Ising magnet in a
ferromagnetic phase (SSB of the $Z_2$ spin symmetry). We perform
the ``decorated domain wall" construction by decorating each Ising
domain wall with a $1d$ Kitaev chain with time-reversal symmetry
$Z_2^T$. The parameters of the Hamiltonian are then tuned to
proliferate these decorated domain walls. When a $1d$ $Z_2$ domain
wall meets (or intersect) with the $1d$ boundary of the system,
the domain wall becomes a $0d$ object decorated with Majorana zero
modes coming from the boundary of the Kitaev's chain. These
Majorana zero modes transform under the time-reversal $Z_2^T$ as
$\gamma_a \rightarrow \gamma_a$, hence any Hermitian fermion
bilinear operator $\ii \gamma_a \gamma_b$ would break the
time-reversal symmetry and hence prohibited. For decoration number
$\nu = 8$, with a proper flavor symmetry between the Majorana
fermion operators $\gamma_a$, the interaction will induce a
many-body symmetric gap between the Majorana modes, and drive the
fermion Green's function at each $0d$ intersection to the
following form~\cite{youzero,kevinQSH}:
\begin{equation}
G_{ab}(\ii\omega) \sim \frac{\ii\omega
\delta_{ab}}{(\ii\omega)^2-m^2}, \label{green}
\end{equation}
in which $m$ is proportional to the strength of the fermion
interaction. There is a uniform gap energy scale in the Green's
function, as long as the eight Majorana fermion operators
$\gamma_a$ form an irreducible representation of the flavor
symmetry, such as a spinor representation of $\SO(7)$ or $\SO(5)$.
$G_{ab}$ approaches zero when $\omega \rightarrow 0$. Notice that
this Green's function takes a different form from a free massive
$0d$ fermion, where a mass term would explicitly break the
time-reversal.

\begin{figure}
\includegraphics[width=0.5\textwidth]{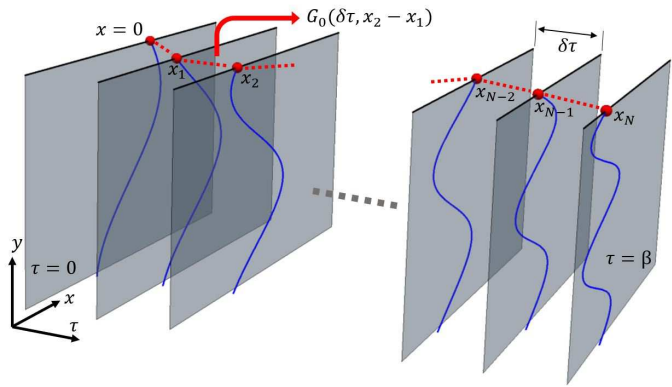}
\caption{Physical picture of the path integral in
Eq.~\ref{pathint}. Here each plane represents the 2d bulk at an
intermediate time, the solid lines are the decorated domain walls,
the circles are gapped Majorana modes at the boundary. The
Majorana modes are connected by dashed lines, which stand for the
local Green's function $G_0(\delta\tau,x_i-x_{i-1})$.}
\label{path3d}
\end{figure}

After being gapped by interaction through the SMG, in the
Euclidean time the Majorana modes (MM) Green's function reads
\beqn G_{ab}(\tau) = G_{\mathrm{MM}}(\tau)\delta_{ab} \sim
\sgn(\tau)e^{-m|\tau|} \delta_{ab}. \label{green2} \eeqn Our goal
is to compute the fermion Green's function after the proliferation
of Ising domain wall. To do this we map the computation of the
Green's function to the following Feynmann path-integral problem
in the $(1+1)d$ space-time, and the different choice of path
$x(\tau)$ physically represents the fluctuation of the domain wall
configuration:
\beqn \label{pathint} G_{ab}(\beta,x) = \delta_{ab} G(\beta, x),
\cr\cr G(\beta, x) \sim \sgn(\beta)\int
D[x(\tau)]\prod_{i=1}^{N}G_0(\delta\tau,\delta x_i)\rho(\delta
\tau, \delta x_i), \eeqn Where $\delta x_i = x_i-x_{i-1}$. Here we
have inserted $N-1$ intermediate steps between the starting point
$(\tau = 0, x_0=0)$ and the end point $(\tau = \beta, x_N = x)$.
$x_i$ is the spatial coordinate along the $1d$ boundary space with
lattice constant $a$ (Fig.~\ref{path3d}).
$\delta\tau=\frac{|\beta|}{N}$ is the time interval for each
intermediate step.

The physical picture behind Eq.~\ref{pathint} is shown in
\figref{path3d}.
$D[x(\tau)] \sim \prod_{i=1}^{N-1}  (dx_i/ a) $ is the integral
measure of Feynmann path integral, which should arise from summing
over $x_i$ along the $1d$ space with lattice constant $a$: $\sum_x
f(x) = \frac{1}{a} \int dx f(x)$. $G_0(\delta\tau,\delta
x)=e^{-m\sqrt{\delta\tau^2+\delta x^2}}$ is the intermediate step
short range propagation of the MM, inherited from
Eq.~\ref{green2}. Here we take the simplest possible form of
$G_0(\delta\tau,\delta x)$ as a generalization of
Eq.~\ref{green2}, which has a Lorentz invariance between $\delta
\tau$ and $\delta x$. $\rho(\delta \tau, \delta x_i)$ with $\delta
x_i = x_i - x_{i-1}$ is an extra factor to control the fluctuation
between intermediate steps, whose form depends on the microscopic
details of domain wall proliferation. We will first consider the
simplest scenario with $\rho(\delta \tau, \delta x_i)=1$.

A direct path integral of Eq.~\ref{pathint} is a bit awkward. We
could change the variables inside each Green's function using the
following trick: \beqn G(\beta,x) &=& \sgn(\beta)\int
D[x(\tau)]\prod_{i=1}^{N} \frac{d\lambda_i
d\phi_i}{2\pi}\prod_{j=1}^{N}G_0(\delta\tau,\phi_j) \cr\cr
&\times&
\exp\left(i\sum_{k=1}^{N}\lambda_k(x_k-x_{k-1}-\phi_k)\right).
\label{trick1} \eeqn Here we introduced two sets of auxiliary
variables: $\lambda_i$ are Lagrangian multipliers; $\phi_i$
substitute the coordinate differences. Now we can integrate out
$x(\tau)$ first, which generates a product of delta functions
$\prod_{i=2}^N\delta(\lambda_i-\lambda_{i-1})$, i.e. all
$\lambda_i$ should equal. Then we can further integrate out
$\lambda_i$, \beqn G(\beta,x) &=& \sgn(\beta)\left( \frac{1}{a}
\right)^{N-1} \int\prod_{i=1}^{N} d\phi_i G_0(\delta\tau,\phi_i)
\cr\cr &\times& \delta(x-\sum_{k=1}^{N}\phi_k). \label{trick2}
\eeqn Now we can perform a Fourier transformation of the spatial
coordinate $x$, and \beqn && G(\beta, k) \sim \frac{1}{a} \int  dx
\ e^{\ii kx} G(\beta,x) \cr\cr &=& \sgn(\beta) \left(\frac{1}{a}
\int d\phi \ e^{\ii k\phi}
G_0(\delta\tau,\phi)\right)^{\frac{|\beta|}{\delta\tau}}.
\label{trick3} \eeqn Here we have replace all $N$ in the
expression by $|\beta|/\delta \tau$, and view $\delta\tau$ as an
independent variable, unrelated to $\beta$. $G(\beta, k)$ takes an
exponential form just like $G_{\mathrm{MM}}(\tau)$, and this
comparison allows us to define an effective mass gap for
$G(\beta,k)$ as follows \beqn
m'(k)&\equiv&-\frac{\ln\left(\frac{1}{a}\int d\phi \ e^{\ii k\phi}
G_0(\delta\tau,\phi)\right)}{\delta\tau} \cr\cr
&=&-\frac{\ln\left(\frac{2m\delta\tau}{a\sqrt{m^2+k^2}}
K_1(\delta\tau\sqrt{m^2+k^2})\right)}{\delta\tau}, \label{mp}
\eeqn so that $G(\beta,k)\sim\sgn(\beta)e^{-m'(k)|\beta|}$.
$K_1(x)$ is the modified Bessel function of the second kind.


With large $m$ or $k$, the effective mass $m'(k)$ is proportional
to $m'(k) \sim \sqrt{m^2 + k^2}$. Hence with large $m, k$ in the
momentum and Matsubara frequency space, the fermion Green's
function takes the approximate form \beqn G(\ii \omega, k) \sim
\frac{\ii \omega}{\omega^2 + m^2 + k^2}. \label{green3} \eeqn This
form of Green's function after SMG is consistent with the fermion
Green's functions obtained in different
models~\cite{kevinQSH,youzero,hepsmg5,smg1,smg2}, after taking the
trace of the Green's functions in these literature, since in our
formalism there is a single component of Majorana fermion in the
Dirac space. As long as $m'(k) > 0$, i.e. the fermion Green's
function decays exponentially in the long time limit, the Fourier
transformation of the Green's function to the Matsubara frequency
space will have zero at $\omega = 0$.

\begin{figure}
\includegraphics[width=0.45\textwidth]{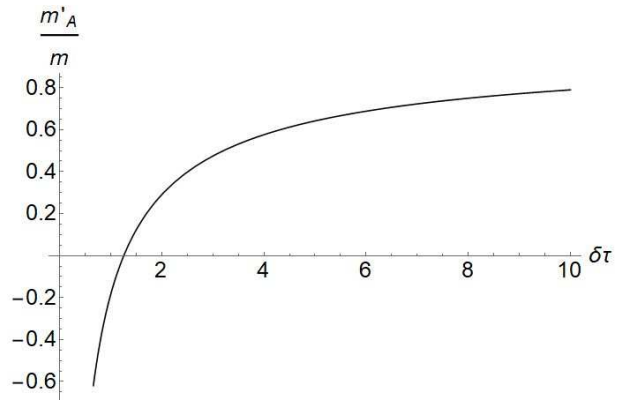}
\caption{Effective mass gap ratio $m'(0)/m$ as a function of
$\delta\tau$. The sign of $m'(0)$ changes from positive to
negative while decreasing $\delta\tau$, suggesting a phase
transition caused by domain wall fluctuation. Here we set
$m=a=1$.} \label{effm}
\end{figure}

The ratio between effective mass gap $m'(0)$ and $m$ as a function
of $\delta\tau$ is shown in \figref{effm}, in which we set
$m=a=1$. When $\delta\tau \ll a$ (meaning there are many intervals
in the path integral), the effective mass $m'$ becomes negative,
this means that in the long time limit the Green's function of the
Majorana fermion no longer exponentially decays. The sign change
of $m'$ signals a phase transition, which is caused by increased
steps of intervals, or physically stronger fluctuation of the
domain wall.

\begin{figure}
\includegraphics[width=0.45\textwidth]{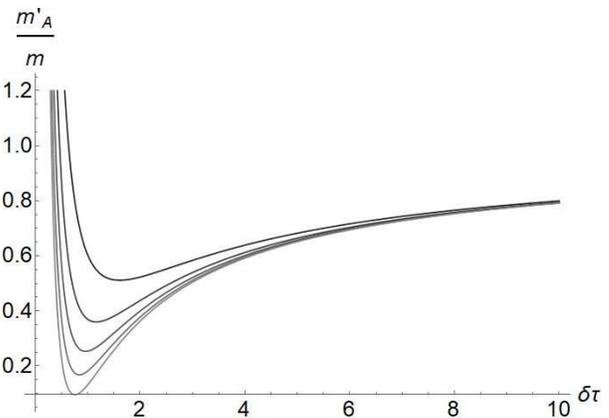}
\caption{Effective mass ratio versus $\delta\tau$ for different
control parameter $A$ with $m=a=1$. From top to bottom
$A=1,2,\cdots 5$. } \label{effmsup}
\end{figure}

Now we turn on the control function $\rho(\delta \tau, \delta
x_i)$ in the Green's function path integral. For example, we can
turn on a Gaussian control function of the fluctuation of the
domain wall: $\rho_A(\delta \tau, \delta x_i) = \exp( - (\delta
x_i / \delta \tau)^2 / A )$, It is not hard to see, from Eqs.
\ref{trick1} to \ref{mp}, that adding such control factor will not
alter the exponential form of the outcome, while the effective
mass gap now reads
\begin{equation}
m'_A=-\frac{\ln\left(\frac{1}{a}\int d\phi \
G_0(\delta\tau,\phi)e^{ - \phi^2 /( A\delta\tau^2) }
\right)}{\delta\tau}. \label{mpsup}
\end{equation}
Results of numerical integral of $\phi$ with different $A$ are
plotted in \figref{effmsup}. Again all $m'_A$ approaches to $m$
when $\delta\tau\to\infty$. And as expected, smaller $A$ will lead
to a larger $m'$, because a smaller $A$ suppresses proliferation
of the domain walls more strongly.

\section{Higher spatial dimensions}


``Decorated defect construction" of TSCs and TIs, or more
generally SPT states can be generalized to higher dimension. As we
mentioned in the introduction, the $3d$ TI in the AIII class can
be constructed by starting with a superfluid with spontaneous
$\U(1)$ symmetry breaking in the $3d$ bulk, then decorate each
vortex line with a Kitaev's chain, and eventually proliferate the
vortex line to restore the $\U(1)$ symmetry in the bulk. A $4d$
TSC with $\SO(3)$ and time-reversal symmetry can be constructed in
a similar way: in the $4d$ space, the hedgehog monopole of a
$\SO(3)$ vector order parameter is a line defect; one can start
with an ordered phase with a $\SO(3)$ vector order parameter, and
decorate the hedgehog monopole line with a $1d$ Kitaev's chain,
and then eventually proliferate the monopole line.

In general one can start with a $d-$dimensional system with
symmetry group $G$ (for example $\SO(d-1)$) that allows one
dimensional topological defect line. In this case each defect line
could be decorated with Kitaev chains, and when these line defects
are proliferated we again presumably obtain gapped TSC with
symmetry $G \times Z_2^T$. When the decorated line defects meet
the $(d-1)-$dimensional boundary, the $0d$ intersection is
decorated with Majorana zero modes. When eight copies of Kitaev's
chains are decorated on the $1d$ defect line, the Majorana modes
at the intersection is gapped by interaction through the SMG
mechanism, and their Green's function is given by Eq.~\ref{green}.

\begin{figure}
\includegraphics[width=0.5\textwidth]{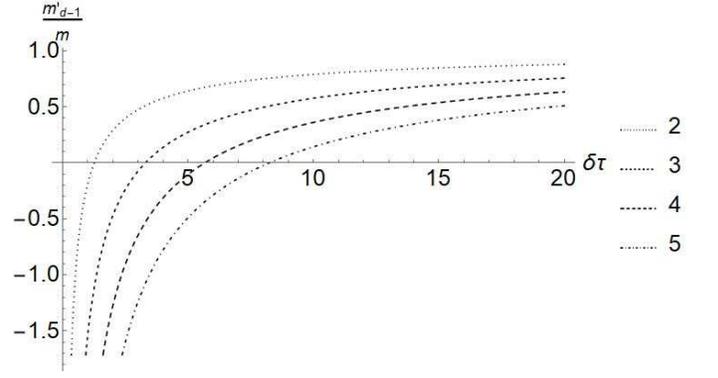}
\caption{$m_{d-1}'/m$ as functions of $\delta\tau$ with $m=a=1$.
From top to bottom $d=2,3,4,5$.} \label{effmd}
\end{figure}

\begin{figure}
\includegraphics[width=0.5\textwidth]{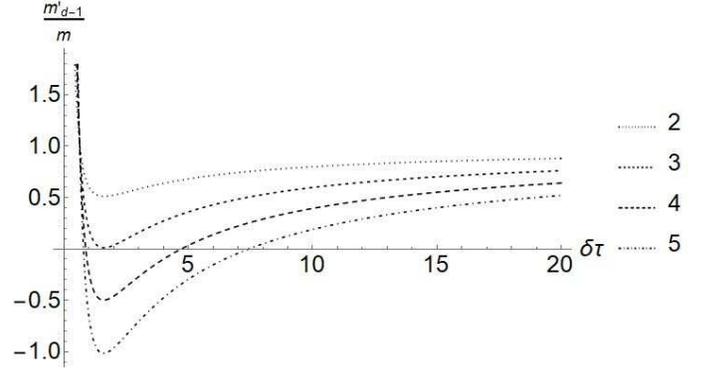}
\caption{$m_{d-1}'/m$ under Gaussian control function as functions
of $\delta\tau$, with $m=a=1$, $d=2$ to $5$ and $A=1$.}
\label{effmdsup}
\end{figure}

Once we proliferate the defects, the fermion Green's function at
the $(d-1)-$dimensional boundary can still reduce to a path
integral problem: \beqn
G^{(d-1)}(\beta,\mathbf{x})&=&\sgn(\beta)\int D[\mathbf{x}(\tau)]
\prod_{i=1}^N G_0^{(d-1)}(\delta\tau, \delta \mathbf{x}_i) \cr \cr
& &\times\rho(\delta \tau, \delta\mathbf{x}_i)\eeqn where $\delta
\mathbf{x}_i = \mathbf{x}_i - \mathbf{x}_{i-1}$, and
$\mathbf{x}_i$ are intermediate positions at the $d-1$ dimensional
boundary (with $\mathbf{x}_0=0$ and $\mathbf{x}_N=\mathbf{x}$) and
$G_0^{(d-1)}(\delta\tau, \delta \mathbf{x}_i)=
\exp\left(-m\sqrt{\delta\tau^2+ \delta \mathbf{x}_i^2}\right)$ is
the simplest extension of Eq.~\ref{green} to $(d-1)-$dimensional
space. The same trick of variable substitution applies here:
\begin{equation}
\begin{aligned}
&G^{(d-1)}(\beta,\mathbf{x})=\sgn(\beta)\int D[\mathbf{x}(\tau)]
\prod_{i=1}^{N}\frac{d\vec{\lambda}_i d\vec{\phi}_i}{(2\pi)^{d-1}}
\\
\times& G_0^{(d-1)}(\delta\tau,\vec{\phi}_i)\rho(\vec{\phi}_i)\exp\left(i\sum_{k=1}^{N}
\vec{\lambda}_k\cdot(\mathbf{x}_k-\mathbf{x}_{k-1}-\vec{\phi}_k)\right).
\end{aligned}
\label{trickd}
\end{equation}
Again, integrating out $\mathbf{x}(t)$ and $\vec{\lambda_i}$
leaves us with a single constraint
$\delta^{(d)}(\mathbf{x}-\sum_{i=1}^{N}\vec{\phi}_i)$. After
integrating over $\mathbf{x}$, we are left with $\int
d\mathbf{x}G^{(d-1)}(\beta,\mathbf{x})
\propto\sgn(\beta)e^{-m_{d-1}'|\beta|}$, in which
\begin{equation}
\begin{aligned}
m'_{d-1}\equiv&-\frac{1}{\delta\tau}\ln\left\{\left(\frac{1}{a}\right)^{d-1}
\int d^{d-1}\vec{\phi} \right. \\
&\qquad\qquad\qquad\qquad\qquad\ \left. \times  G_0^{(d-1)}(\delta\tau,\vec{\phi})\rho(\vec{\phi})\right\} \\
=&-\frac{\ln\left(\frac{ma}{\pi}\left(\frac{2\pi\delta\tau}{ma^2}\right)^{\frac{d}{2}}K_{d/2}(m\delta\tau)\right)}{\delta\tau}(\text{for
}\rho(\vec{\phi})=1)
\end{aligned}
\end{equation}
is the new effective mass gap. The ratios between $m_{d-1}'$ and
$m$ is plotted in \figref{effmd}, and we can see from the plot
that increasing spatial dimension makes the effective mass gap
smaller, indicating that fluctuation is stronger for higher
dimensions. Indeed, for higher dimensions there is more space for
the proliferation of path $\mathbf{x}(\tau)$. When the Gaussian
control function $\rho_A(\delta \tau, \delta \mathbf{x}_i) =
\exp(- (\delta \mathbf{x}_i / \delta \tau)^2 / A )$ is turned on,
the stronger fluctuation for higher dimensions makes the Gaussian
suppression less effective (see \figref{effmdsup}). Nevertheless,
for nonzero $A$, $m'_{d-1}$ can still be positive (and hence there
is a zero in the Green's function) for a broad range of
parameters.

\section{Scenarios without Translation symmetry}

One of the previous observations and arguments for the existence
of fermion Green's function zero, is based on the quantized
topological number for TSC and TI defined with the fermion Green's
function~\cite{gurarie1,gurarie2}. A typical topological number
can be defined in the Matsubara frequency and momentum space of
the Euclidean space-time fermion Green's
function~\cite{volovik1,volovikbook,wanggreen1,wanggreen2,wanggreen3}:
$n \sim \int d\omega d^dk \ \mathrm{tr} [ B (G^{-1}
\partial G) \wedge (G^{-1} \partial G) \cdots ] $, where $G$ is
the matrix of the fermion Green's function, and $B$ is a matrix in
the flavor space. The number $n$ must be a quantized integer
mathematically, and it can only change when the Green's function
has singularity.

The number $n$ can change through two types of ``transitions". The
first type of transition is a physical transition where
$G^{-1}(\ii \omega = 0)$ vanishing to zero at certain momentum,
i.e. the fermions become gapless. In this case the physical
topological transition coincides with the transition of the
topological number. However, one can easily notice that in the
definition of $n$, $G^{-1}$ and $G$ are on an equal footing, hence
theoretically the topological number mentioned above can also
change when $G(\ii \omega = 0) = 0$, i.e. when the Green's
function has a zero. Hence when the TSC or TI is trivialized by
the interaction, although there is no unavoidable phase transition
between the TSC (or TI) and a trivial insulator, the topological
number $n$ still has to change discontinuously somewhere in the
phase diagram, and since there is no real physical transition, the
number $n$ has to change through zero of the Green's function.

This argument for Green's function zeros relies on the quantized
topological number in the momentum space, hence it requires the
translation symmetry. But none of the TSC and TI in the ``ten-fold
way" classification requires translation symmetry, hence it is
natural to ask whether the Green's function zeros persist when the
translation symmetry is broken. Normally the translation symmetry
breaking is caused by disorder, i.e. a random potential energy.
But a fermion bilinear potential term $\ii \gamma_a \gamma_b$
breaks the time-reversal symmetry of the decorated Kitaev's chain.
Hence the most natural translation symmetry breaking perturbation
that can be turned on in the system, is a spatial dependent random
four-fermion interaction, $i.e. $ a randomized $m(x)$ in
Eq.~\ref{green2}, Eq.~\ref{pathint}.

Now Eq.~\ref{pathint} is modified to \beqn && G(\beta, x) =
\sgn(\beta)\int D[x(\tau)] \cr\cr && e^{ \sum_{i = 1}^N - m(x_i)
\sqrt{\delta \tau^2 + \delta x^2_i} } \ \rho_A (\delta \tau,
\delta x_i). \eeqn $m(x_i)$ is a space-dependent but
time-independent function. In principle $m(x_i)$ could be any
function of space. Here we focus on the situation when $m(x_i) = m
+ \delta m(x_i)$, where $m$ is a positive constant, while $\delta
m (x_i)$ is random function of space with zero mean and Gaussian
distribution. After disorder average, the expression for the
Green's function is \beqn && \overline{G(\beta)} = \sgn(\beta)\int
D[x(\tau)] \cr\cr && e^{ \sum_{i = 1}^N - m \sqrt{\delta \tau^2 +
\delta x^2_i} + \sum_{j, k} \Delta \delta(x_j - x_k) \sqrt{\delta
\tau^2 + \delta x^2_j} \sqrt{\delta \tau^2 + \delta x^2_k} }
\cr\cr && \times \rho_A (\delta \tau, \delta x_i). \eeqn $\Delta$
is given by the Gaussian distribution of $\delta m(x_i)$: \beqn
\overline{\delta m(x_j) \ \delta m(x_k)} \sim \Delta \delta(x_j -
x_k). \eeqn

\begin{figure}
\includegraphics[width=0.5\textwidth]{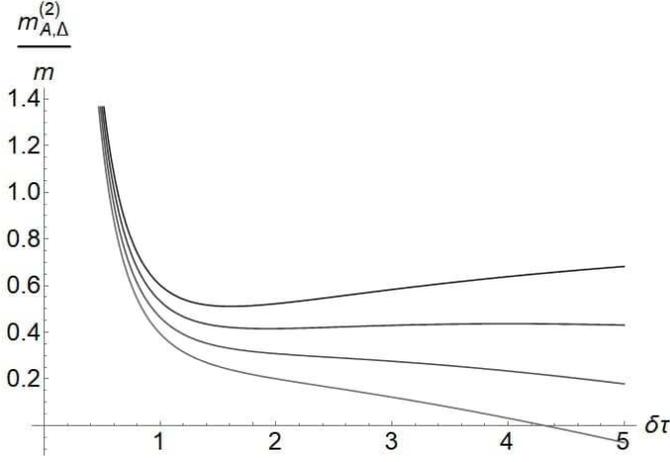}
\caption{Effective mass gaps $m^{(2)}_{A, \Delta}$ for the lowest
order of disorder averaged Green's function
$\overline{G(\beta)}_0$ as functions of $\delta\tau$. Here we set
$m=a=A=1$, $\Delta=0$, $0.05$, $0.1$, $0.15$ from the top to
bottom.} \label{mDelta}
\end{figure}

The delta function $\delta(x_j - x_k)$ is only nonzero when $x_j =
x_k$. This condition automatically satisfies when $j = k$, but may
still happen when $j \neq k$, meaning the path $x(\tau)$ returns
to the same spatial location at different time instances. We first
consider the contribution from disorder average when $ j = k$:
\beqn \overline{G(\beta)}_0 &=& \sgn(\beta)\int D[x(\tau)] e^{
\sum_{i = 1}^N - m \sqrt{\delta \tau^2 + \delta x^2_i} + \sum_{i}
\Delta (\delta \tau^2 + \delta x^2_i) } \cr\cr &\times& \rho_A
(\delta \tau, \delta x_i). \cr\cr &=&
\sgn(\beta)\exp\left(-\beta(m'_{\tilde{A}}-\Delta\delta\tau)\right).\eeqn
Here $m'_{\tilde{A}}$ with $\tilde{A} = \frac{A}{1-A\Delta}$ is
the effective mass gap defined in Eq.~\ref{mpsup} with a new
Gaussian control parameter $\tilde{A} = \frac{A}{1-A\Delta}$. Thus
$\overline{G(\beta)}_0$ behaves identically to the previously
computed Gaussian suppressed Green's function, albeit with a
slower decaying rate, $m^{(2)}_{A,\Delta} \equiv
m'_{\tilde{A}}-\Delta\delta\tau$. The effect of disorder on the
effective mass $m^{(2)}_{A,\Delta}$ is plotted in \figref{mDelta}.

For the contribution from $j\neq k$ (we assume that $j < k$
hereafter), we can expand the Green's function into powers of
$\Delta$. The first order of this expansion is \beqn
\overline{G(\beta)}_1 &\equiv&
\Delta\sgn(\beta)e^{\beta\Delta\delta\tau}\sum_{j\neq k}\int
D[x(\tau)] e^{ \sum_{i = 1}^N - m \sqrt{\delta \tau^2 + \delta
x^2_i} } \cr\cr &\times& \sum_{j\neq k}
\delta(x_j-x_k)\sqrt{\delta \tau^2 + \delta x^2_j} \sqrt{\delta
\tau^2 + \delta x^2_k}\cr\cr &\times& \prod_{i=1}^N
\rho_{\tilde{A}} (\delta \tau, \delta x_i).\eeqn Again using the
previous variable substitution, we obtain \beqn
\overline{G(\beta)}_1&=&\Delta\sgn(\beta)e^{\beta\Delta\delta\tau}\sum_{j\neq
k}\int D[x(\tau)]\prod_{i=1}^{N} \frac{d\lambda_i
d\phi_i}{2\pi}\cr\cr &\times&
e^{-\sum_{i}m\sqrt{\delta\tau^2+\phi_i^2}-\ii\lambda_i(x_i-x_{i-1}-\phi_i)}
\delta(x_j-x_k)\cr\cr &\times&
\sqrt{\delta\tau^2+\phi_j^2}\sqrt{\delta\tau^2+\phi_k^2} \
\prod_{i=1}^N \rho_{\tilde{A}} (\delta \tau, \phi_i). \eeqn
Integrating over all $x_i$ and $x_0$, we obtain the product of a
series of delta functions \beqn &
&\prod_{i=1}^{j-1}\delta(\lambda_i-\lambda_{i+1})\prod_{i=j+1}^{k-1}
\delta(\lambda_i-\lambda_{i+1})\prod_{i=k+1}^{N-1}
\delta(\lambda_i-\lambda_{i+1})\cr\cr &\times&\delta(\lambda_N)
\delta(\lambda_k-\lambda_{k+1}+\lambda_j-\lambda_{j+1}). \eeqn For
example $\delta(\lambda_N)$ comes from $\int dx_N$. Integrating
out other $x_i$ will enforce $\lambda_{i}=0$ for all $i>k$;
$\lambda_{i}=\lambda_1$ for all $i\leq j$; and
$\lambda_i=\lambda_k$ for all $j<i\leq k$. The final delta
function above thus also enforces $\lambda_i=\lambda_1 = 0$ with
all $i \leq j$. Notice that $\lambda_k$ is unconstrained here,
because $\delta(x_j-x_k)$ effectively removes one $\delta$
function constraint for $\lambda_i$'s. So we are left with the a
single integral of $\lambda_k\equiv\lambda$, and the result
is\beqn
\overline{G(\beta)}_1&=&2\Delta\sgn(\beta)e^{\beta\Delta\delta\tau}
\sum_{j<k}(\tilde{G}_\Delta(0))^{N-(k-j)-1}\partial_m
\tilde{G}_\Delta(0)\cr\cr &\times& \int d\lambda
(\tilde{G}_\Delta(\lambda))^{k-j-1}\partial_m
\tilde{G}_\Delta(\lambda)\cr\cr &=& \overline{G(\beta)}_0\times
2\Delta\frac{\partial_m\tilde{G}_\Delta(0)}{\tilde{G}_\Delta(0)}
\sum_{h = 1}^{N - 1}(N - h)\cr\cr &\times& \int d\lambda
\left(\frac{\tilde{G}_\Delta(\lambda)}{\tilde{G}_\Delta(0)}\right)^{h-1}
\frac{\partial_m\tilde{G}_\Delta(\lambda)}{\tilde{G}_\Delta(0)}.
\label{greendisorder1} \eeqn Here $\tilde{G}_\Delta(\lambda)=
\int\frac{d\phi}{a}e^{\ii\lambda\phi-m\sqrt{\delta\tau^2+\phi^2}}
\rho_{\tilde{A}} (\delta \tau, \phi)$.

\begin{figure}
\includegraphics[width=0.5\textwidth]{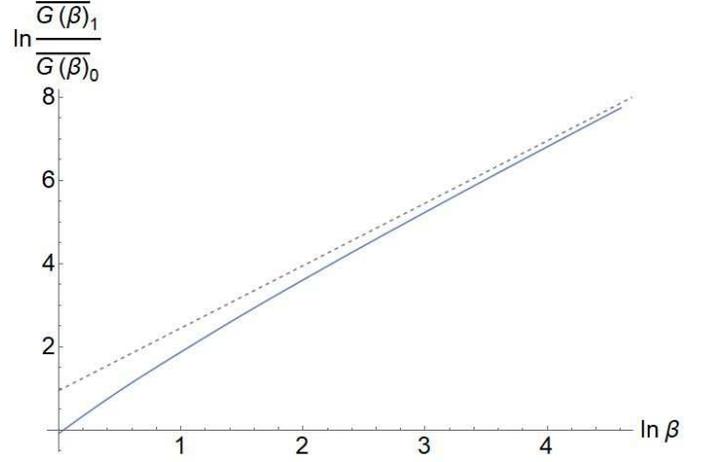}
\caption{Log-log plot of numerical integration of $
\overline{G(\beta)}_1/\overline{G(\beta)}_0$. Here we set
$m=a=A=1$, $\delta\tau=0.3$ and $\Delta=0.1$. The dashed line is a
guide to the eye with slope $3/2$.} \label{g/g}
\end{figure}

Numerical integration of $\lambda$ in the expression above shows
that the ratio between the first two orders of the $\Delta$
expansion, i.e. $ \overline{G(\beta)}_1/\overline{G(\beta)}_0$
approaches $\beta^{3/2}$ for large $\beta$ (see \figref{g/g}).
Thus at large $\beta$ the overall behavior of the first order term
in the $\Delta$ expansion still exponentially decays with $\beta$.
The behavior of large $\beta$ can be understood in the following
way: The integral of $\tilde{G}_\Delta(\lambda)$ can be
approximated by replacing $\sqrt{\delta \tau^2 + \phi^2}$ by
$\delta \tau + |\phi|$ in the exponent, which means
$\frac{\tilde{G}_\Delta(\lambda)}{\tilde{G}_\Delta(0)}$ and
$\frac{\partial_m\tilde{G}_\Delta(\lambda)}{\tilde{G}_\Delta(0)}$
behaves approximately as $e^{-\frac{A}{4(1-A\Delta)}\lambda^2}$.
The $\lambda$ integral gives a $\frac{1}{\sqrt{h}}$ factor in each
term of the summation of Eq.~\ref{greendisorder1}. Eventually
$\overline{G(\beta)}_1/\overline{G(\beta)}_0$ is evaluated as
\begin{equation}
\sum_{h = 1}^{N}\frac{N - h}{\sqrt{h}}\sim \int_0^\beta dx
\frac{\beta-x}{\sqrt{x}} \sim \beta^{3/2}.
\end{equation} And as long as the overall behavior of
$\overline{G(\beta)}$ decays exponentially with $\beta$, the
Fourier transformation of $\overline{G(\beta)}$ has a zero at
$\omega = 0$.

At higher dimensions, it is less likely for $\mathbf{x}_j =
\mathbf{x}_k$ at $j \neq k$, i.e. it is less likely for a path
$\mathbf{x}(\tau)$ to return to exactly the same spatial location
at two different time instances. Hence we expect that for higher
spatial dimensions the zeroth order $\overline{G(\beta)}_0$ in the
formulation above should be even more accurate.

\section{The ``avoided" topological transition in the bulk}

As we discussed in the introduction, besides the nontrivial
boundary state, there is another prominent feature of a TI and
TSC: there must be an unavoidable bulk topological transition
between the TI or TSC and the trivial insulator when tuning the
parameter of the bulk Hamiltonian. However, once the TI or TSC is
trivialized by interaction, not only can the boundary state be
trivially gapped, the bulk topological transition also becomes
avoidable~\cite{fidkowski1,fidkowski2}: there is an adiabatic path
in the phase diagram connecting the original TI (or TSC) phase and
the original trivial insulator phase without closing the gap at
all. In this case the original topological transition is also
called ``unnecessary transition"~\cite{bisenthil,jianavoid}. In
fact there is a close relation between the boundary state and the
bulk topological transition. The simplest model to visualize such
relation is the Chalker-Coddington model~\cite{cc,cc2}, which was
first developed for the integer quantum Hall transition. This
boundary-bulk relation can be made much more general for strongly
interacting symmetry protected topological
states~\cite{bulkboundary}. In general the bulk topological
transition between the trivial phase and the SPT phase can be
viewed as growing islands of the SPT phase inside a trivial phase,
and the unavoidable topological transition originates from the
nontrivial interface states between the trivial and SPT phases.
When the interfaces percolate, the bulk is at the topological
transition. Using this picture, our real-space calculation for
Green's function in the previous sections for a $d-$dimensional
boundary, also applies to the avoided bulk topological transition
at $d-$dimensions.

{\it --- Summary}

In this work we demonstrate the existence of the Green's function
zero as a remaining trace of nontrivial topology, after the system
acquires a fully gapped spectrum after the mechanism of symmetric
mass generation. Our method mostly relies on the real space
decorated defect construction of the SPT states, and it does not
require spatial symmetries such as translation.

This work is supported by NSF Grant No. DMR-1920434, and the
Simons Foundation.

\bibliography{zero}

\end{document}